\documentclass[letterpaper, 10 pt, twocolumn]{ieeeconf}

\IEEEoverridecommandlockouts                              
\usepackage{amsbsy}

\usepackage{amsmath}
\usepackage{amssymb}
\usepackage{times}
\usepackage{graphicx}
\usepackage{xspace}
\usepackage{paralist} 
\usepackage{setspace} 
\usepackage{xypic}
\xyoption{curve}
\usepackage{latexsym}
\usepackage{theorem}
\usepackage{ifthen}
\usepackage{subfigure}
\usepackage{turnstile}

{\theoremheaderfont{\it} \theorembodyfont{\rmfamily}
\newtheorem{theorem}{Theorem}
\newtheorem{lemma}[theorem]{Lemma}
\newtheorem{definition}[theorem]{Definition}

\newtheorem{proposition}[theorem]{Proposition}
\newtheorem{remark}[theorem]{Remark}

}

\renewcommand{\baselinestretch}{1}

\long\def\symbolfootnote[#1]#2{\begingroup%
\def\thefootnote{\fnsymbol{footnote}}\footnote[#1]{#2}\endgroup}

\begin{document}

\title{\LARGE \bf A Random Dynamical Systems Approach to Filtering in Large-scale Networks}
\author{Soummya Kar, Bruno Sinopoli, and Jos\'e M.~F.~Moura
\thanks{
The authors are with the Department of Electrical and Computer Engineering, Carnegie Mellon University, 5000 Forbes Ave, Pittsburgh, PA 15213. {\tt\small soummyak@andrew.cmu.edu, \{brunos, moura\}@ece.cmu.edu }
}
}

\maketitle
\thispagestyle{empty}
\pagestyle{empty}

\begin{abstract}The paper studies the problem of filtering a discrete-time linear
system observed by a network of sensors. The sensors share a
common communication medium to the estimator and transmission is
bit and power budgeted. Under the assumption of conditional
Gaussianity of the signal process at the estimator (which may be
ensured by observation packet acknowledgements), the conditional
prediction error covariance of the optimum mean-squared error
filter is shown to evolve according to a random dynamical system
(RDS) on the space of non-negative definite matrices. Our RDS
formalism does not depend on the particular medium access protocol
(randomized) and, under a minimal distributed observability
assumption, we show that the sequence of random conditional
prediction error covariance matrices converges in distribution to a
unique invariant distribution (independent of the initial filter
state), i.e., the conditional error process is shown to be
ergodic. Under broad assumptions on the medium access protocol, we
show that the conditional error covariance sequence satisfies a
Markov-Feller property, leading to an explicit characterization of
the support of its invariant measure. The methodology adopted in this
work is sufficiently general to envision this application to sample
path analysis of more general hybrid or switched systems, where
existing analysis is mostly moment-based.
\end{abstract}


\section{Introduction}
\label{introduction}

\subsection{Background and Motivation}
\label{backmot}
Networked Control Systems (NCS) have been proposed as the paradigm to model, design and analyze control systems where the effects of computation and communication on the performance of the closed loop system cannot be neglected and need to be incorporated in the model. NCS are amenable to describe large-scale systems where components may be spatially distributed and demand the services of a communication network to exchange information. Integration of communication models traditionally renders the problem of design and analysis very complex, as the traditional mathematical machinery employed in control systems in general is not adequate to include  in its framework stochastic models of communication.

A new approach capable of capturing the complexity of NCS by being able to extract fundamental limitations and relevant properties is needed. In~\cite{Riccati-weakconv} we proposed to use the Random Dynamical Systems~\cite{ArnoldChueshovbook} framework to characterize the problem of Kalman filtering where observations may be dropped in the communication between the sensors and the estimator, as in the case where a wireless sensor network is employed to observe the evolution of a particular dynamical system. The problem was originally analyzed in~\cite{Bruno}. In that work the authors proved the existence of a critical packet arrival rate, below which the estimation error covariance, a stochastic variable dependent upon the realization of the packet arrival process, would have infinite mean, rendering the estimate useless.

In this case the optimal estimator would fail to track the system. The model proposed in~\cite{Bruno} has been widely adopted and
extended by several
authors~\cite{Liu:04,Gupta:05,Xu:05,Minyi:07,kp-fb:07j,Craig:07,xx07}.
Although many present extensions to general Markov chains and
account for smart sensors sending local estimates instead of
observations, all the results are established with respect to mean
stability, i.e., boundedness of the mean covariance. This metric
is unsatisfactory in many applications, as it does not provide
information about the fluctuations of the error covariance that
could grow and be unusable for long time intervals. We would like
to characterize the asymptotic behavior of its distribution--the
goal of this paper. In~\cite{Riccati-weakconv}, the authors showed how RDS can be used to fully characterize the problem, by providing conditions on the convergence in distribution of the error covariance to a stationary distribution whose support can also be explicitly characterized.

In this paper we extend the result of ~\cite{Riccati-weakconv} to the case of many sensors. In particular it is not our goal to evaluate the performance of the filter with respect to a specific sensor scheduling policy or communication protocol, but to uncover the macroscopic properties of such policies and their effect on the performance of the overall systems. Under fairly general assumptions on the sensor schedule (which may be a system design objective or imposed by the random communication medium) and a minimal detectability condition, we show ergodicity of the conditional error covariance sequence and explicitly characterize the support of the resulting invariant distribution. For clarity we focus on temporally i.i.d. schedules in this paper, but note how the approach can be extended to cover more general Markovian scheduling policies. We believe that this paradigm can be extended to the verification of properties of stochastic switched or hybrid systems where nowadays the analysis tends to be moment based.

We briefly describe the organization of the rest of the paper. The problem is rigorously formulated in Section~\ref{prob_form} and a description of the main results of the paper appear in Section~\ref{main_res}. Section~\ref{RDS_form} presents the RDS formulation of the error covariance evolution. Several properties of the resulting RDS are presented in Section~\ref{sec:prop_RDS}, whereas, Section~\ref{main_proof} provides outlines for the proofs. Finally, Section~\ref{conclusions} concludes the paper.

\subsection{Notation}
\label{notprel}
Denote by: $\mathbb{R}$, the reals; $\mathbb{R}^{M}$,
the $M$-dimensional Euclidean space; $\mathbb{T}$, the integers; $\mathbb{T}_{+}$, the non-negative integers; $\mathbb{N}$, the natural numbers; and $\mathcal{X}$, a generic space. The separable Banach
space of symmetric $M\times M$ matrices is denoted by $\mathbb{S}^{M}$,
equipped with the induced 2-norm. The subset $\mathbb{S}^{M}_{+}$
of positive semidefinite matrices is a closed, convex, solid,
normal, minihedral cone in $\mathbb{S}^{n}$, with non-empty
interior $\mathbb{S}^{M}_{++}$, the set of positive definite
matrices. The cone $\mathbb{S}_{+}^{M}$ induces a partial order in $\mathbb{S}^{M}$, namely, for $X,Y\in \mathbb{S}^{M}$, we write $X\preceq Y$, if $Y-X\in \mathbb{S}^{M}_{+}$.
In case $X\preceq Y$ and $X\neq Y$, we write $X\prec Y$. We also have a strong order by virtue of the non-emptiness of the interior of $\mathbb{S}_{+}^{M}$, the set of positive definite matrices $\mathbb{S}_{++}^{M}$, where we write $X\ll Y$, if
$Y-X\in\mathbb{S}_{++}^{M}$.

\section{Problem Formulation}
\label{prob_form}

\subsection{System Model}
\label{sys_model_label} We consider a discrete-time linear dynamical system being observed by a network of $N$ sensors. The signal model is given by:
\begin{equation}
\label{sys_model}
\mathbf{x}(t+1)=A\mathbf{x}(t)+\mathbf{w}(t)
\end{equation}
where $\mathbf{x}(t)\in\mathbb{R}^{M}$ is the signal (state) vector with initial state $\mathbf{x}(0)$ being distributed as a zero mean Gaussian vector with covariance $P_{0}$ and the system noise $\{\mathbf{w}(t)\}$ is an uncorrelated zero mean Gaussian sequence independent of $\mathbf{x}(0)$ with covariance $Q$. The observation at the $n$-th sensor $\mathbf{y}_{n}(t)\in\mathbb{R}^{m_{n}}$ at time $t$ is of the form:
\begin{equation}
\label{obs_n}
\mathbf{y}_{n}(t)=C_{n}\mathbf{x}(t)+\mathbf{v}_{n}(t)
\end{equation}
where $C_{n}\in\mathbb{R}^{m_{n}\times M}$ and $\{\mathbf{v}_{n}(t)\}$ is an uncorrelated zero mean Gaussian observation noise sequence with covariance $P_{n}\gg \mathbf{0}$. Also, the noise sequences at different sensors are independent of each other, the system noise process and the initial system state. Because of limited capability of the sensors, typically the dimension of $\mathbf{y}_{n}(t)$ is much smaller than that of the signal process and the observation process at each sensor is not sufficient to make the pair $\{\mathbf{x}(t),\mathbf{y}_{n}(t)\}$ observable. Thus the sensors need to collaborate and to achieve this, they share a common wireless medium to an estimator (possibly remote.) Such a medium is bit and power constrained, the channel access is opportunistic and hence at every iteration\footnote{Iteration refers to the discrete time index $\mathbb{T}_{+}$.} only a subset of the $N$ sensors are able to send their observations to the estimator. We assume there exists a \emph{randomized sensor schedule}, where at each time $t$ only a subset of sensors, randomly chosen, get channel access and successfully send their observations to the estimator. We formalize this as follows:
\begin{definition}[Sensor Schedule]:
\label{sensor_schedule}
Let $\mathfrak{P}$ be the power set of $\{1,2,\cdots,n,\cdots,N\}$, i.e., the set of $2^{N}$ of its subsets, including the null set. We number the elements of $\mathfrak{P}$ as $i$, where $i$ ranges from 0 to $2^{N}-1$ and w.l.o.g. assume that 0 corresponds to the null set. A sensor schedule (randomized) $\mathcal{D}$ is a probability distribution on the set $\mathfrak{P}$, such that, at time $t$, the set $\mathcal{I}(t)$ of transmitting sensors is chosen randomly according to the distribution $\mathcal{D}$ from $\mathfrak{P}$. Also, we assume that the random process $\{\mathcal{I}(t)\}_{t\geq 0}$ is an i.i.d. sequence (taking values in the set $\mathfrak{P}$ of subsets of $\{1,\cdots,N\}$),\footnote{Since we identify every element of $\mathfrak{P}$ with an unique integer $i\in\{0,\cdots,2^{N}-1\}$, for all purposes, we take $\mathcal{I}(t)$ to assume values in $\{0,\cdots,2^{N}-1\}$.} i.e., the set of transmitting sensors at time $t$ is independent of the assignment at all previous times, leading to a memoryless channel.

To every schedule $\mathcal{D}$ we assign a probability vector $\lambda^{\mathcal{D}}=[\lambda_{0}^{\mathcal{D}},\cdots,\lambda_{2^{N}-1}^{\mathcal{D}}]^{T}\in[0,1]^{2^{N}}$, such that, $\sum_{i=0}^{2^{N}-1}\lambda_{i}^{\mathcal{D}}=1$ and
\begin{equation}
\label{probD}
\mathbb{P}\left[\mathcal{I}(t)=i~|~I(s),~~0\leq s<t\right]=\lambda_{i}^{\mathcal{D}},~~~\forall t\in\mathbb{T}_{+}
\end{equation}
\end{definition}
The schedule defined above leads to temporally independent sensor assignments, but allows correlated transmissions at each time $t$ among the sensors. The model is thus fairly general, capturing the class of memoryless (temporally) transmission schemes and subsumes many existing scheduling policies in networked control systems, for example,~\cite{Bruno}. The model is not applicable to channels with memory, however, with the framework developed in this paper we can cope with a large class of Markovian channels as explained later. For clarity of presentation, we assume the memoryless scheduling policy $\mathcal{D}$ defined above.

Suppose a particular $\mathcal{D}$ is in place. To a generic element $i$ of $\mathfrak{P}$, we assign the corresponding subset $i_{S}\subset\{1,2,\cdots,N\}$
indicating the sensors contained in $i$. We denote $i_{S}$ by $
i_{S}=\{i_{1},\cdots,i_{|i_{S}|}\}$, where $|i_{S}|$ denotes the cardinality of $i_{S}$ and $i_{j}$ for $j=1,\cdots,|i_{S}|$ denotes the $j$-th sensor in the set $i_{S}$. With this notation in place, for every $i\in\mathfrak{P}$, we define
\begin{equation}
\label{cumC} C^{i}=[C_{i_{1}}^{T},C_{i_{2}}^{T},\cdots,C_{i_{|i_{S}|}}^{T}]^{T}
\end{equation}
\begin{equation}
\label{cumC2} \mathbf{v}^{i}(t)=[\mathbf{v}_{i_{1}}^{T}(t),\cdots,\mathbf{v}_{i_{|i_{S}|}}^{T}(t)]^{T},~~~\forall t\in\mathbb{T}_{+}
\end{equation}
and let $R^{i}$ denote the covariance of the zero mean Gaussian vector $\mathbf{v}^{i}(t)$, which we assume to be positive definite. Then, at time $t$, if some non-trivial subset of sensors reports to the estimator, i.e., if $\mathcal{I}(t)\neq\phi$, the cumulative observation $\mathbf{y}^{\mathcal{I}(t)}(t)$ arriving at the estimator is given by
\begin{equation}
\label{cumC3}
\mathbf{y}^{\mathcal{I}(t)}(t)=C^{\mathcal{I}(t)}\mathbf{x}(t)+\mathbf{v}^{\mathcal{I}(t)}(t)
\end{equation}
We assume that, alongside the observation, every sensor sends an acknowledgement to the estimator whenever it gets to transmit. Thus the information available to the estimator at time $t$ is given by
\begin{equation}
\label{cumC4}
\mathfrak{I}(t)=\{\mathbf{y}^{\mathcal{I}(s)}(s),\mathcal{I}(s),~~0\leq s <t\}
\end{equation}
and the objective is to find the minimum mean squared estimator (mmse), which is the conditional mean
\begin{equation}
\label{cumC5}
\widehat{\mathbf{x}}(t|t)=
\mathbb{E}\left[\mathbf{x}(t)~|~\mathfrak{I}(t)\right]
\end{equation}
Conditioned on the information sequence $\{\mathfrak{I}(t)\}$, the signal process becomes conditionally Gaussian, and the conditional mean can be recursively updated by a modified (time-varying) Kalman filter (\cite{Liptser-Shiryaev}.) It is sufficient to study the evolution of the one-step predictor
\begin{equation}
\label{cumC5}
\widehat{\mathbf{x}}(t|t-1)=
\mathbb{E}\left[\mathbf{x}(t)~|~\mathfrak{I}(t-1)\right]
\end{equation}
and the conditional prediction error covariance
{\small
\begin{equation}
\label{cumC6}
P(t)=\mathbb{E}[\left(\mathbf{x}(t)-\widehat{\mathbf{x}}(t|t-1)\right)\left(\mathbf{x}(t)-\widehat{\mathbf{x}}(t|t-1)\right)^{T}~|~\mathcal{I}(t-1)]
\end{equation}
}
It can be shown that the random sequence $\{P(t)\}_{t\in\mathbb{T}_{+}}$ evolves according to a random algebraic Riccati equation (RARE) as:
{\small
\begin{equation}
\label{cumC7}
P(t+1)=\left\{ \begin{array}{l}
                    AP(t)A^{T}+Q~~~~~~~~~~~~~~~~~\mbox{\textbf{if $\mathcal{I}(t)=0$}} \\ \\
                    AP(t)A^{T}+Q-AP(t)(C^{\mathcal{I}(t)})^{T}\left(C^{\mathcal{I}(t)}P(C^{\mathcal{I}(t)})^{T}\right.\\
                    \left.+R^{\mathcal{I}(t)}\right)^{-1}C^{\mathcal{I}(t)}P(t)A^{T}~~~~~~~\mbox{\textbf{otherwise}}
                   \end{array}
          \right.
\end{equation}}
with initial condition $P(0)=P_{0}$.
Under the assumption of memoryless scheduling, the sequence $\{P(t)\}$ is a Markov process whose asymptotic properties are of interest to us. To this end we define the continuous functions $f_{i}:\mathbb{S}_{+}^{M}\longmapsto\mathbb{S}_{+}^{M}$ for $i=0,\cdots,2^{N}-1$ by
{\small
\begin{equation}
\label{def_f0}
f_{0}(X)=AXA^{T}+Q
\end{equation}
\begin{equation}
\label{def_fs}
f_{i}(X)=AXA^{T}+Q-AX(C^{i})^{T}\left(C^{i}X(C^{i})^{T}+R^{i}\right)^{-1}C^{i}XA^{T}
\end{equation}
}
for $i=1,\cdots,2^{N}-1$.
With the above notation, the evolution of the RARE sequence $\{P(t)\}$ can be described by:
\begin{equation}
\label{evP}
P(t+1)=f_{\mathcal{I}(t)}(P(t)),~~~P(0)=P_{0}
\end{equation}
Since the sequence $\{f_{\mathcal{I}(t)}\}$ comprises of i.i.d. random maps, the process $\{P(t)\}$ is Markov. The RARE described above can be viewed as a generalization of the RARE studied in~\cite{Riccati-weakconv}, where the sequence of prediction error covariance matrices involved switching between two random functions, one of which was the Lyapunov update like $f_{0}$ and the other was the Riccati update like the $f_{i}, i=1,\cdots, 2^{N}-1$ considered in eqn.~(\ref{def_fs}). Hence, as will be noted later, several technical results in this paper are direct generalizations of their specific cases considered in~\cite{Riccati-weakconv}.

As is the case with even deterministic single sensor systems, some form of stabilizability and detectability is needed to guarantee stability of the filtering error process. We consider the following notion of weak detectability in a networked system:
\begin{definition}[Weak Detectability]:
\label{weak_det} A sensor schedule $\mathcal{D}$ is called \emph{weakly detectable} if there exists $i\neq 0$ in $\mathfrak{P}$ with $\lambda_{i}^{\mathcal{D}}>0$, such that the pair $(C^{i},A)$ is detectable.
\end{definition}
It is to be noted that the term `weak' in the above definition not only indicates that the above is a weak condition on the sensor network, but also has implications to weak convergence (convergence in distribution) of the sequence $\{P(t)\}$ as will be justified later. Note that weak detectability defined above does not require the pair $(C_{n},A)$ to be detectable for any sensor $n$, which may be too strong a condition, especially when the signal process is of large dimension and the individual sensors have limited observation capabilities. However, if the network is large, it is fair to assume that a subset of sensors exists, whose cumulative observations lead to a detectable system, and with some arbitrarily small but positive probability all sensors in this subset can transmit simultaneously. So far the concept of weak detectability, as stated above, has been an abstract notion and we are yet to show its effect on the long term behavior of the RARE sequence.

We also make the following assumption on the signal process:

\textbf{Assumption E.1}: The pair $(A,Q^{1/2})$ is stabilizable, $A$ is unstable and $Q\gg 0$.

\subsection{Stability notions and scheduling protocols}
\label{stabnot}In the sequel, when considering the sequence $\{P(t)\}$, we denote the probability and expectation operators by $\mathbb{P}^{\mathcal{D},P_{0}}\left[\cdot\right]$ and $\mathbb{E}^{\mathcal{D},P_{0}}\left[\cdot\right]$ respectively to emphasize the dependence on the scheduling policy and the initial condition. Also, the corresponding sequence of measures induced by the RARE process on $\mathbb{S}_{+}^{M}$ is denoted by $\{\mathbb{\mu}^{\mathcal{D},P_{0}}\}_{t\in\mathbb{T}_{+}}$.

We study the following notions of stability of the RARE sequence:
\begin{definition}[Stochastic boundedness]:
\label{stoch_bound_def} A scheduling policy $\mathcal{D}$ is said to achieve stochastic boundedness for the RARE sequence, if for all initial conditions $P_{0}\in\mathbb{S}_{+}^{M}$, the sequence $\{P(t)\}$ stays stochastically bounded, i.e., for all $P_{0}\in\mathbb{S}_{+}^{M}$
\begin{equation}
\label{stoch_bound_eqn}
\lim_{K\rightarrow\infty}\sup_{t\in\mathbb{T}_{+}}\mathbb{P}^{\mathcal{D},
P_{0}}\left(\left\|P_{t}\right\|>K\right)=0
\end{equation}
\end{definition}
\begin{definition}[Moment stability]
\label{mom-stab} A scheduling policy $\mathcal{D}$ is said to stabilize the RARE sequence in the $k$-th moment ($1\leq k<\infty$) if, for all initial conditions $P_{0}\in\mathbb{S}_{+}^{M}$, the sequence $\{P(t)\}$ has bounded moments of order $k$, i.e., for all $P_{0}\in\mathbb{S}_{+}^{M}$
\begin{equation}
\label{mom_stab_eqn}
\sup_{t\in\mathbb{T}_{+}}\mathbb{E}^{\mathcal{D},
P_{0}}\left[\left\|P_{t}\right\|^{k}\right]<\infty
\end{equation}
\end{definition}
We often abuse terminology to call a scheduling policy stochastically bounded (s.b.) or $k$-th moment stable.

The following proposition relates the two stability notions
and establishes the important connection between weak detectability and stability:
\begin{proposition}
\label{prop_crel} Let the pair $(A,Q^{1/2})$ be stabilizable and $A$ unstable. We then have the following:
\begin{itemize}
\item[i] Moment stability of any order implies stochastic boundedness, i.e., if a scheduling policy $\mathcal{D}$ is moment stable of order $k$ for some $1\leq k<\infty$, then it is stochastically bounded.

\item[ii] Weak detectability implies stochastic boundedness, i.e., a weakly detectable scheduling policy $\mathcal{D}$ is s.b.
\end{itemize}
\end{proposition}
\begin{proof} The first part is a general property of stochastic sequences and the proof uses Chebyshev type of inequalities to relate probability to expectation whose details can be found in~\cite{Riccati-weakconv}. The proof of the second part is more involved and lengthy, but an immediate generalization of its specific case involving a single sensor with intermittent transmissions is studied in~\cite{Riccati-weakconv},\cite{Riccati-moddev} and is omitted.
\end{proof}
\begin{remark}Proposition~\ref{prop_crel} shows that stochastic boundedness is weaker than moment stability and whereas weak detectability is sufficient to ensure s.b., stronger detectability assumptions are required to conclude moment stability. This is illustrated by an example of a single sensor system as described below. Our interest in stochastic boundedness comes from the fact that it is sufficient to ensure ergodicity of the process $\{P(t)\}$, the key point of the paper. The rest of the paper concerns this important relation between stochastic boundedness (and hence weak detectability by Proposition~\ref{prop_crel}) and ergodicity of the RARE sequence, and this is achieved by using tools from the theory of RDS.
\end{remark}
\textbf{Example: Single sensor system with intermittent transmission}
\\
For a single sensor system, a scheduling policy corresponds to assigning a probability $\overline{\gamma}\in [0,1]$ of packet transmission. This is studied in detail in~\cite{Riccati-weakconv},\cite{Riccati-moddev}. We assume that $\overline{\gamma}>0$ and the pair $(A,C_{1})$ is observable, which corresponds to the notion of weak detectability in the general case. In~\cite{Riccati-weakconv},\cite{Riccati-moddev} it was shown that $\overline{\gamma}>0$ is a necessary and sufficient condition for stochastic boundedness if $A$ is unstable (the case of invertible $C$ was shown in~\cite{Riccati-weakconv}, whereas~\cite{Riccati-moddev} proves the general observable case), whereas mean stability of the sequence $\{P(t)\}$ requires $\overline{\gamma}$ to be greater than a critical value which increases to 1 as $A$ becomes more and more unstable (see~\cite{Bruno} for a lower bound on this critical probability.) Moreover, in~\cite{Riccati-weakconv} it was shown that stochastic boundedness ($\overline{\gamma}>0$) is sufficient to ensure ergodicity of the RARE process and hence we can establish that the sequence $\{P(t)\}$ converges to a unique invariant distribution even if it is not stable in the mean. In this paper, we take this further and show that a weakly detectable schedule is sufficient to ensure ergodicity of $\{P(t)\}$, even if it is not moment stable of any order $k$.

\section{Main Results: Invariant Distribution}
\label{main_res} We state the main results of the paper. Proof outlines are provided in Section~\ref{main_proof}, the details will appear elsewhere. Several key technical components of the proofs are direct generalizations of the development in~\cite{Riccati-weakconv} for the single sensor case.

The first result concerns the ergodicity of the conditional error covariance process.
\begin{theorem}
\label{main_th} Let Assumption~\textbf{E.1} hold and the scheduling policy $\mathcal{D}$ be weakly detectable. Then
there exists a unique invariant distribution
$\mathbb{\mu}^{\mathcal{D}}$ supported on the set of positive definite matrices $\mathbb{S}_{++}^{M}$, s.t.~the RARE sequence
$\left\{P_{t}\right\}_{t\in\mathbb{T}_{+}}$ (or the sequence
$\left\{\mathbb{\mu}_{t}^{\mathcal{D},P_{0}}\right\}_{t\in\mathbb{T}_{+}}$
of measures) converges weakly to
$\mathbb{\mu}^{\mathcal{D}}$ from any initial condition
$P_{0}$.

\end{theorem}

The second result explicitly determines the support of the
invariant measure $\mathbb{\mu}^{\mathcal{D}}$.
\begin{theorem}
\label{supp_inv}Let the hypotheses of Theorem~\ref{main_th} hold. Define the set $\mathcal{J}(\mathcal{D})\subset\mathfrak{P}$ as:
\begin{equation}
\label{def_J}
\mathcal{J}(\mathcal{D})=\{i\in\mathfrak{P}~|~\lambda_{i}^{\mathcal{D}}>0~\mbox{and the pair $(C^{i},A)$ is detectable}\}
\end{equation}
For every $i\in\mathcal{J}(\mathcal{D})$ define the set\footnote{Below, in definition of $\mathcal{S}_{i}$, $s$ can take the
value 0, implying $P^{\ast}_{i}\in\mathcal{S}$.}
\begin{eqnarray}
\label{def_S}
\mathcal{S}_{i}=\left\{f_{l_{1}}\circ
f_{l_{2}}\circ\cdots\circ
f_{l_{s}}\left(P^{\ast}_{i}\right)~|~l_{r}\in\{0,\cdots,2^{N}-1\},\right.\nonumber \\ \left.1\leq r\leq
s,\:s\in\mathbb{T}_{+}\right\}
\end{eqnarray}
where $P_{i}^{\ast}$ is the unique fixed point of the operator $f_{i}:\mathbb{S}_{+}^{M}\longmapsto\mathbb{S}_{+}^{M}$.\footnote{The stabilizability of $(A,Q^{1/2})$ and the detectability of $(C^{i},A)$ guarantee the existence of such a $P_{i}^{\ast}$.} We then have:
\begin{itemize}
\item[i] $\mathcal{S}_{i}=\mathcal{S}_{j}$ for all $i,j\in\mathcal{J}(\mathcal{D})$.\\
\item[ii] The invariant measure $\mathbb{\mu}^{\mathcal{D}}$ is supported on the closure of $\mathcal{S}_{i}$, where $i\in\mathcal{J}(\mathcal{D})$, i.e.,
    \begin{equation}
    \label{def_support}
    \mbox{supp}\left(\mathbb{\mu}^{\mathcal{D}}\right)=\mbox{cl}\left(\mathcal{S}_{i}\right),~~i\in\mathcal{J}(\mathcal{D})
    \end{equation}
    where $\mbox{cl}$ denotes the topological closure of a set on the space $\mathbb{S}_{+}^{M}$.
\end{itemize}
\end{theorem}

\textbf{Discussions:} Theorem~\ref{main_th} establishes the ergodicity of the prediction error covariance sequence $\{P(t)\}$ under the assumption that the scheduling policy $\mathcal{D}$ is weakly detectable. The support of the invariant distribution $\mathbb{\mu}^{\mathcal{D}}$ is explicitly characterized in Theorem~\ref{supp_inv} in terms of the network parameters. Note that, in particular, if $\mathcal{D}_{1}$ and $\mathcal{D}_{2}$ are two weakly detectable scheduling policies, such that there exists $i\in\mathfrak{P}$ with $\lambda_{i}^{\mathcal{D}_{1}},\lambda_{i}^{\mathcal{D}_{2}}>0$ and $(C^{i},A)$ is detectable, then
\begin{equation}
\label{disc1}
\mbox{supp}\left(\mathbb{\mu}^{\mathcal{D}_{1}}\right)=\mbox{supp}\left(\mathbb{\mu}^{\mathcal{D}_{2}}\right)=\mbox{cl}\left(\mathcal{S}_{i}\right)
\end{equation}
where $\mathcal{S}_{i}$ is defined in eqn.~(\ref{def_S}). As shown in the single sensor case (\cite{Riccati-weakconv}), the invariant distributions $\mathbb{\mu}^{\mathcal{D}}$ are generally not absolutely continuous w.r.t. the Lebesgue measure and the supports $\mbox{supp}\left(\mathbb{\mu}^{\mathcal{D}}\right)$ are highly fractured subsets of $\mathbb{S}_{++}^{M}$.

As is noted in the proof of Theorem~\ref{main_th} (Subsection~\ref{proof_main_th}), the only requirement is stochastic boundedness of the sequence $\{P(t)\}$ from every initial condition. Weak detectability ensures this condition (Proposition~\ref{prop_crel}) and is generally stronger than the requirement on stochastic boundedness. Thus, the conclusions of Theorem~\ref{main_th} may stay valid under even weaker conditions on the schedule $\mathcal{D}$, provided we are able to establish stochastic boundedness of $\{P(t)\}$ from every initial condition. However, in the absence of weak detectability, an explicit characterization of the invariant distribution, as in Theorem~\ref{supp_inv} may not be possible.

Similarly, we can extend Theorem~\ref{main_th} to stationary temporally Markovian schedules (i.e., the selection process $\{\mathcal{I}(t)\}$ is a stationary Markov process rather than i.i.d.). The proof and conclusions of Theorem~\ref{main_th} will be unchanged, as the RDS formulation of the error process requires only stationarity (see~\cite{Chueshov}.) However, with a non i.i.d. selection process $\{\mathcal{I}(t)\}$, the sequence $\{P(t)\}$ no longer stays Markov and hence the assertions of Theorem~\ref{supp_inv} may not be valid.

\section{Random Dynamical System Formulation}
\label{RDS_form} In this section, we formulate the RARE process as an RDS evolving on $\mathbb{S}_{+}^{M}$. An excellent treatment of the theory of RDS can be found in~\cite{ArnoldChueshov,Chueshov} and the concepts relevant to us are detailed in~\cite{Riccati-weakconv}. To minimize overlap with~\cite{Riccati-weakconv} and due to lack of space, we mention RDS facts, as and when required.

We start by defining a random dynamical system
(RDS). We follow the notation
in~\cite{ArnoldChueshov,Chueshov}.
\begin{definition}[RDS]\label{defn_RDS} A RDS with (one-sided) time
$\mathbb{T}_{+}$ and state space $\mathcal{X}$ is the pair
$(\theta,\varphi)$: 
\begin{itemize}
\item[\textbf{A)}] A metric dynamical system
$\theta=\left(\Omega,\mathcal{F},\mathbb{P},
\left\{\theta_{t},t\in\mathbb{T}\right\}\right)$
with two-sided time $\mathbb{T}$, i.e., a probability space
$(\Omega,\mathcal{F},\mathbb{P})$ with a family of transformations
$\{\theta_{t}:\Omega\longmapsto\Omega\}_{t\in\mathbb{T}}$ such
that
\begin{itemize}
\item[\textbf{A.1)}]$\theta_{0}=id_{\Omega},
\:\,\theta_{t}\circ\theta_{s}=\theta_{t+s},\:\,\forall
t,s\in\mathbb{T}$
\item[\textbf{A.2)}]$(t,\omega)\longmapsto\theta_{t}\omega$ is
measurable.
\item[\textbf{A.3)}]$\theta_{t}\mathbb{P}=\mathbb{P}\:\forall
t\in\mathbb{T}$, i.e., $\mathbb{P}\left(\theta_{t}B\right)=\mathbb{P}(B)$ for all $B\in\mathcal{F}$ and all $t\in\mathbb{T}$.
\end{itemize}
\item[\textbf{B)}] A cocycle $\varphi$ over $\theta$ of continuous
mappings of $\mathcal{X}$ with time $\mathbb{T}_{+}$, i.e., a
measurable mapping
\begin{equation}
\label{def_RDS} \varphi:\mathbb{T}_{+}\times\Omega\times
\mathcal{X}\rightarrow \mathcal{X}, \:(t,\omega,X)\longmapsto\varphi(t,\omega,X)
\end{equation}
\begin{itemize}
\item[\textbf{B.1)}] The mapping
$X\longmapsto\varphi(t,\omega,X)\equiv\varphi(t,\omega)X$ is
continuous in $X$  $\forall\,t\in\mathbb{T}_{+},\,\omega\in\Omega$.
\item[\textbf{B.2)}] The mappings
$\varphi(t,\omega)\doteq\varphi(t,\omega,\cdot)$ satisfy the
cocycle property: $\:\:\forall\,
t,s\in\mathbb{T}_{+},\,\,\omega\in\Omega$,
\begin{equation}
\label{def_RDS1}
\varphi(0,\omega)=id_{\mathcal{X}},\:\:\varphi(t+s,\omega)=
\varphi(t,\theta_{s}\omega)\circ\varphi(s,\omega)
\end{equation}
\end{itemize}
\end{itemize}
\end{definition}
 In a RDS, randomness is captured by the  space
$(\Omega,\mathcal{F},\mathbb{P})$. Iterates indexed by $\omega$
indicate pathwise construction. For example, if $X_{0}$ is the
deterministic  state 
 at $t=0$, the state at $t\in\mathbb{T}_{+}$ is 
\begin{equation}
\label{def_RDS2} X_{t}(\omega)=\varphi\left(t,\omega,X_{0}\right)
\end{equation}
The measurability assumptions 
  guarantee that the state $X_{t}$ is a well-defined random variable.
Also,  the iterates are defined for non-negative
(one-sided) time; however, the family of transformations
$\left\{\theta_{t}\right\}$ is two-sided, which is purely for technical
convenience, as will be seen later.

We now show that the sequence $\left\{P_{t}\right\}$ generated by the RARE can be modeled as the sequence of iterates (in
the sense of distributional equivalence) of a suitably defined
RDS.

Fix a schedule $\mathcal{D}$ and define:
$\left(\widetilde{\Omega},\widetilde{\mathcal{F}},
\widetilde{\mathbb{P}}^{\mathcal{D}}\right)$,
where $\widetilde{\Omega}=\mathfrak{P}$,
$\widetilde{\mathcal{F}}$ is the power set of $\mathfrak{P}$ and the discrete probability measure $\widetilde{\mathbb{P}}^{\mathcal{D}}$ is defined by
\begin{equation}
\label{prob_tilde}
\widetilde{\mathbb{P}}^{\mathcal{D}}\left[\{i\}\right]=\lambda_{i}^{\mathcal{D}},~~~\forall i\in\mathfrak{P}
\end{equation}
Also, define the product space,
$\left(\Omega,\mathcal{F},\mathbb{P}^{\mathcal{D}}\right)$, where
$\Omega=\times_{t\in\mathbb{T}}\widetilde{\Omega}$ and
$\mathcal{F}$ and $\mathbb{P}^{\mathcal{D}}$ are the
 product $\sigma$-algebra and the product
measure\footnote{Note the difference between the measure
$\mathbb{P}^{\mathcal{D}}$ and the measures
$\mathbb{P}^{\mathcal{D},P_{0}}$ defined in
Subsection~\ref{sys_model_label}.}. From the
construction, a sample point $\omega\in\Omega$ is a
two-sided sequence taking values in the discrete set $\mathfrak{P}=\{0,\cdots,2^{N}-1\}$ and, since
$\mathbb{P}^{\mathcal{D}}$ is the product of
$\widetilde{\mathbb{P}}^{\mathcal{D}}$, the projections are
i.i.d. random variables with common distribution $\mathcal{D}$. Define the family of transformations
$\left\{\theta^{R}_{t}\right\}_{t\in\mathbb{T}}$ on $\Omega$ as the family of
left-shifts
\begin{equation}
\label{def_RDS3} \theta_{t}^{R}\omega=\omega(t+\cdot),\:\forall
t\in\mathbb{T}
\end{equation}
With this, the space
$\left(\Omega,\mathcal{F},\mathbb{P}^{\overline{\gamma}},
\left\{\theta_{t}^{R},t\in\mathbb{T}\right\}\right)$
is the canonical path space of a two-sided stationary (in
fact, i.i.d.) sequence equipped with the left-shift operator;
hence, (e.g., \cite{Kallenberg}) it satisfies the
Assumptions~\textbf{A.1)-A.3)} to be a metric dynamical system;
in fact, it is ergodic.

Recall the functions $f_{i}(X)$ in eqns.~(\ref{def_f0},\ref{def_fs}).
Define the function
$\widetilde{f}:\Omega\times\mathbb{S}_{+}^{M}\longmapsto\mathbb{S}_{+}^{M}$
by
\begin{equation}
\label{def_RDS4} \widetilde{f}(\omega,X)=f_{\omega(0)}(X)
\end{equation}
 Since the
projection map from $\omega$ to $\omega(0)$ is measurable
(continuous) and $f_{i}(X)$ is jointly measurable in $i,X$,
 it follows that $\widetilde{f}(\cdot)$ is
jointly measurable in $\omega,X$. Define the function
$\varphi^{R}:\mathbb{T}_{+}\times\Omega\times\mathbb{S}_{+}
\longmapsto\mathbb{S}_{+}$
by
\begin{eqnarray}
\label{def_RDS5} \varphi^{R}(0,\omega,X)&=&X\\
\label{RDS6}
\varphi^{R}(1,\omega,X)&=&\widetilde{f}(\omega,X)\\
\label{def_RDS7}
\varphi^{R}(t,\omega,X)&=&\widetilde{f}
\left(\theta^{R}_{t-1}\omega,\varphi^{R}(n-1,\omega,X)\right)
\end{eqnarray}
It follows from the measurability of the transformations
$\left\{\theta_{t}^{R}\right\}$, the measurability of $\widetilde{f}(\cdot)$,
and the fact that $\mathbb{T}_{+}$ is countable that the function
$\varphi^{R}(t,\omega,X)$ is jointly measurable in $t,\omega,X$.
Finally, $\varphi^{R}(\cdot)$ defined above satisfies
Assumption~\textbf{B.1)} by the continuity of the $f_{i}$s, and
Assumption~\textbf{B.2)} follows by the construction given by
eqns.~(\ref{def_RDS5}-\ref{def_RDS7}). Thus, the pair
$\left(\theta^{R},\varphi^{R}\right)$ is an RDS over $\mathbb{S}_{+}^{M}$.
Given a deterministic initial condition
$P_{0}\in\mathbb{S}_{+}^{M}$, it follows that the sequence
$\left\{P_{t}\right\}_{t\in\mathbb{T}_{+}}$ generated by the RARE
eqn.~(\ref{cumC7}) is equivalent in the sense of distribution
to the sequence
$\left\{\varphi^{R}(t,\omega,P_{0})\right\}_{t\in\mathbb{T}_{+}}$ generated
by the iterates of the above constructed RDS, i.e.,
\begin{equation}
\label{def_RDS100}
P_{t}\ndtstile{}{d}\varphi^{R}(t,\omega,P_{0}),\:\forall
t\in\mathbb{T}_{+}
\end{equation}
Thus, investigating the distributional properties of
$\left\{P_{t}\right\}_{t\in\mathbb{T}_{+}}$ is equivalent to analyzing the
distributional properties of
$\left\{\varphi^{R}(t,\omega,P_{0})\right\}_{t\in\mathbb{T}_{+}}$, which we
carry out in the rest of the paper.

In the sequel, we use the pair $(\theta,\varphi)$ to denote a
generic RDS and $\left(\theta^{R},\varphi^{R}\right)$ for the one constructed
above for the RARE.

\section{Properties of $\left(\theta^{R},\varphi^{R}\right)$}
\label{prop_RDS}

\label{sec:prop_RDS}

\subsection{Facts about generic RDS}
\label{facts_genRDS} We review concepts on RDS
(see~\cite{ArnoldChueshov,Chueshov} for details.)
 Consider a generic RDS $(\theta,\varphi)$ with state space
$\mathcal{X}$ as in Definition~\ref{defn_RDS}. Let $\mathcal{X}=V_{+}$, where $V_{+}$ is a
closed, convex, solid, normal, minihedral cone of a real Banach space $V$. Denote by $\preceq$ the
partial order induced by $V_{+}$ in $\mathcal{X}$ and $<<$ denotes
the corresponding strong order. For clarity the reader may assume in the following that $V=\mathbb{S}^{M}$, $\mathcal{X}=V_{+}=\mathbb{S}_{+}^{M}$ and $\preceq, <<$ represent the order induced by positive definiteness in the space $\mathbb{S}_{+}^{M}$. In fact, the RDS $\left(\theta^{R},\varphi^{R}\right)$ considered in this paper evolves in the cone $\mathbb{S}_{+}^{M}$ of positive semidefinite matrices, where the partial order is induced by positive definiteness.

\begin{definition}[Order-Preserving RDS]:
\label{Order} A RDS $(\theta,\varphi)$ with state space $V_{+}$
is order-preserving if $\:\forall
t\in\mathbb{T}_{+},\:\omega\in\Omega,\:X,Y\in V_{+}$,
\begin{equation}
X\preceq
Y\:\Longrightarrow\:\varphi(t,\omega,X)\preceq\varphi(t,\omega,Y)
\end{equation}
\end{definition}

\begin{definition}[Sublinearity]\label{sublinearity} An order-preserving RDS $(\theta,\varphi)$ with state space $V_{+}$ is
sublinear if for every $X\in V_{+}$ and $\lambda\in (0,1)$
we have
\begin{equation}
\label{prop_RDS}
\lambda\varphi(t,\omega,X)\preceq\varphi(t,\omega,\lambda
X),\:\forall t>0,\:\omega\in\Omega
\end{equation}
The RDS is strongly sublinear if in addition to
eqn.~(\ref{prop_RDS}), we have
\begin{equation}
\label{prop_RDS2}
\lambda\varphi(t,\omega,X)\ll\varphi(t,\omega,\lambda X),\:\forall
t>0,\:\omega\in\Omega,\:X\in\mbox{int}\,V_{+}
\end{equation}
\end{definition}

\begin{definition}[Equilibrium]\label{equilibrium} A random
variable $u:\Omega\longmapsto V_{+}$ is called an equilibrium
(fixed point, stationary solution) of the RDS $(\theta,\varphi)$
if it is invariant under $\varphi$, i.e.,
\begin{equation}
\label{prop_RDS3}
\varphi\left(t,\omega,u(\omega)\right)=u\left(\theta_{t}\omega\right),\:\forall
t\in\mathbb{T}_{+},\:\omega\in\Omega
\end{equation}
If eqn.~(\ref{prop_RDS3}) holds $\forall\,\omega\in\Omega$,
except on  set of $\mathbb{P}$ measure zero, $u$ is an
almost equilibrium.
\end{definition}
Since the transformations $\left\{\theta_{t}\right\}$ are
measure-preserving, i.e.,
$\theta_{t}\mathbb{P}=\mathbb{P},\:\forall t$, we have
\begin{equation}
\label{prop_RDS4} u\left(\theta_{t}\omega\right)\ndtstile{}{d}
u(\omega),\:\forall t
\end{equation}
By eqn.~(\ref{prop_RDS3}), for an
almost equilibrium $u$, the iterates in the sequence
$\left\{\varphi\left(t,\omega,u(\omega)\right)\right\}_{t\in\mathbb{T}_{+}}$
have the same distribution, which is the distribution of $u$.

%

\begin{definition}[Orbit]\label{orbit} For a random variable $u:\Omega\longmapsto
V_{+}$, we define the \emph{forward} orbit $\eta^{f}_{u}(\omega)$
emanating from $u(\omega)$ as the random set
$\left\{\varphi\left(t,\omega,u(\omega)\right)\right\}_{t\in\mathbb{T}_{+}}$.
The forward orbit gives the sequence of iterates of the RDS
starting at $u$.

Although $\eta^{f}_{u}$ is the object of  interest,
for technical convenience (as will be seen later), we also define the
\emph{pull-back} orbit $\eta^{b}_{u}(\omega)$ emanating from $u$
as the random set
$\left\{\varphi\left(t,\theta_{-t}\omega,u\left(\theta_{-t}\omega\right)
\right)\right\}_{t\in\mathbb{T}_{+}}$.
\end{definition}
We establish asymptotic properties for
the pull-back orbit $\eta^{b}_{u}$. This is because it is more convenient and because analyzing $\eta_{u}^{b}$ leads to understanding the asymptotic distributional properties for $\eta^{f}_{u}$. In fact, the random sequences
$\left\{\varphi\left(t,\omega,u(\omega)\right)\right\}_{t\in\mathbb{T}_{+}}$
and $\left\{\varphi\left(t,\theta_{-t}\omega,u\left(\theta_{-t}\omega\right)
\right)\right\}_{t\in\mathbb{T}_{+}}$ are equivalent in distribution. In other words,
\begin{equation}
\label{orbit1}
\varphi\left(t,\omega,u(\omega)\right)\ndtstile{}{d}\varphi
\left(t,\theta_{-t}\omega,u\left(\theta_{-t}\omega\right)\right),\:\forall
t\in\mathbb{T}_{+}
\end{equation}
This follows from
$\theta_{t}\mathbb{P}=\mathbb{P},\:\forall t\in\mathbb{T}$. Thus,
in particular, we have the following assertion.
\begin{lemma}
\label{orbit_lemma} Let the sequence
$\left\{\varphi\left(t,\theta_{-t}\omega,u\left(\theta_{-t}\omega\right)
\right)\right\}_{t\in\mathbb{T}_{+}}$
converge in distribution to a measure $\mu$ on $V_{+}$, where
$u:\Omega\longmapsto V_{+}$ is a random variable. Then the
sequence
$\left\{\varphi\left(t,\omega,u(\omega)\right)\right\}_{t\in\mathbb{T}_{+}}$
also converges in distribution to the measure $\mu$.
\end{lemma}

We now introduce some notions of boundedness of RDS, which will be
used in the sequel.

\begin{definition}[Boundedness]\label{boundedness} Let
$a:\Omega\longmapsto V_{+}$ be a random variable. The pull-back
orbit $\eta_{a}^{b}(\omega)$ emanating from $a$ is
bounded on $U\in\mathcal{F}$ if there exists a random variable $C$
on $U$ s.t.
\begin{equation}
\label{boundedness1}
\left\|\varphi\left(t,\theta_{-t}\omega,a\left(\theta_{-t}\omega\right)
\right)\right\|\leq
C(\omega),\:\forall t\in\mathbb{T}_{+},~\omega\in U
\end{equation}
\end{definition}

\begin{definition}[Conditionally Compact RDS]\label{cond_comp} An
RDS $(\theta,\varphi)$ in $V_{+}$ is conditionally
compact if for any $U\in\mathcal{F}$ and pull-back orbit
$\eta_{a}^{b}(\omega)$ that is bounded on $U$ there exists a
family of compact sets $\{K(\omega)\}_{\omega\in U}$ s.t.
\begin{equation}
\label{cond_comp1}
\lim_{t\rightarrow\infty}\mbox{dist}\left(\varphi\left(t,\theta_{-t}\omega,
a\left(\theta_{-t}\omega\right)\right),K(\omega)\right)=0,\:\omega\in
U
\end{equation}
\end{definition}
It is to be noted that conditionally compact is a topological
property of the space $V_{+}$. In particular, an RDS in a finite
dimensional space $V_{+}$ is conditionally compact.

We now state a limit set dichotomy result for a class of
sublinear, order-preserving RDS.
\begin{theorem}[Corollary 4.3.1. in~\cite{Chueshov}]\label{LSD}
Let $V$ be a separable Banach space with a normal solid cone
$V_{+}$. Assume that $(\theta,\varphi)$ is a strongly sublinear
conditionally compact order-preserving RDS over an ergodic metric
dynamical system $\theta$. Suppose that $\varphi(t,\omega,0)\gg 0$
for all $t>0$ and $\omega\in\Omega$. Then precisely one of the
following applies:
\begin{itemize}
\item[\textbf{(a)}] For any $X\in V_{+}$ we have
\begin{equation}
\label{LSD1}
\mathbb{P}\left(\lim_{t\rightarrow\infty}\left\|\varphi\left(t,\theta_{-t}\omega,X\right)\right\|=\infty\right)=1
\end{equation}
\item[\textbf{(b)}] There exists a unique almost equilibrium
$u(\omega)\gg 0$ defined on a $\theta$-invariant set\footnote{A
set $A\in\mathcal{F}$ is called $\theta$-invariant if
$\theta_{t}A=A$ for all $t\in\mathbb{T}$.}
$\Omega^{\ast}\in\mathcal{F}$ with
$\mathbb{P}\left(\Omega^{\ast}\right)=1$ such that, for any random
variable $v(\omega)$ possessing the property $0\preceq
v(\omega)\preceq\alpha u(\omega)$ for all $\omega\in\Omega^{\ast}$
and deterministic $\alpha>0$, the following holds:
\begin{equation}
\label{LSD2}
\lim_{t\rightarrow\infty}\varphi\left(t,\theta_{-t}\omega,v\left(\theta_{-t}\omega\right)\right)=u(\omega),\:\omega\in\Omega^{\ast}
\end{equation}
\end{itemize}
\end{theorem}

\label{prop_Ric_RDS} In this subsection, we establish some
properties of the RDS $\left(\theta^{R},\varphi^{R}\right)$ modeling the
RARE.

\begin{lemma}
\label{prop_R} The RDS $\left(\theta^{R},\varphi^{R}\right)$ with state space
$\mathbb{S}_{+}^{M}$ is order-preserving. In other words, $\:\forall
t\in\mathbb{T}_{+},~\omega\in\Omega,~X,Y\in\mathbb{S}_{+}^{M}$,
\begin{equation}
\label{prop_R1} X\preceq
Y~\Longrightarrow~\varphi^{R}(t,\omega,X)\preceq\varphi^{R}(t,\omega,Y)
\end{equation}
Also, if $Q$ is positive definite, i.e., $Q\gg 0$, it is strongly
sublinear.
\end{lemma}
\begin{proof} The proof uses properties of the functions $f_{0}$ (Lyapunov) and $f_{i}$s (Riccati) and is routine given the arguments in~\cite{Riccati-weakconv} for the single sensor case. We omit it.
\end{proof}

\section{Theorems~\ref{main_th},\ref{supp_inv}: Proof Outlines}
\label{main_proof} We outline the proofs of Theorems~\ref{main_th},\ref{supp_inv}. The proof of Theorem~\ref{main_th} relies on the RDS formulation of the RARE, and we highlight the significant steps in Subsection~\ref{proof_main_th}, the left-out details being a generalization of the single-sensor case in~\cite{Riccati-weakconv} are omitted.

The proof of Theorem~\ref{supp_inv} rests on establishing the RARE sequence as a Markov Feller process and expressing the support of the invariant distribution as the intersection of the topological lower limits of all possible orbits. Due to space constraints, we do not provide details in this paper and rather explain intuitively how it parallels the corresponding development in~\cite{Riccati-weakconv}.

\subsection{Proof of Theorem~\ref{main_th}}
\label{proof_main_th}



\begin{lemma}
\label{step} Consider the RDS $\left(\theta^{R},\varphi^{R}\right)$. Let Assumption~\textbf{E.1} hold and the scheduling policy $\mathcal{D}$ is such that the RARE sequence $\{P(t)\}$ remains stochastically bounded for every initial condition $P_{0}\in\mathbb{S}_{+}^{M}$ (by Lemma~\ref{prop_crel} this condition is satisfied, in particular, when $\mathcal{D}$ is weakly detectable.) Then, there exists a unique almost equilibrium
$u^{\mathcal{D}}(\omega)\gg 0$ defined on a
$\theta^{R}$-invariant set $\Omega^{\ast}\in\mathcal{F}$ with
$\mathbb{P}^{\mathcal{D}}\left(\Omega^{\ast}\right)=1$ s.t.~for any random variable $v(\omega)$ possessing the property
$0\preceq v(\omega)\preceq\alpha u^{\mathcal{D}}(\omega) \,\forall\, \omega\in\Omega^{\ast}$ and deterministic $\alpha>0$, the
following holds:
\begin{equation}
\label{step1}
\lim_{t\rightarrow\infty}\varphi^{R}
\left(t,\theta^{R}_{-t}\omega,v\left(\theta^{R}_{-t}\omega\right)
\right)=u^{\overline{\gamma}}(\omega),\:\omega\in\Omega^{\ast}
\end{equation}
\end{lemma}

\begin{proof}
From Lemma~\ref{prop_R}, $\left(\theta^{R},\varphi^{R}\right)$ is
strongly sublinear  and order-preserving. It is conditionally
compact because the space $\mathbb{S}_{+}^{M}$ is finite
dimensional. Also, the cone $\mathbb{S}_{+}^{M}$ satisfies the
conditions required in the hypothesis of Theorem~\ref{LSD}. From
the properties of the functions $f_{i}$, we note for $t>0$
\begin{equation}
\label{step2} \varphi^{R}\left(t,\omega,0\right)
=
f_{\omega(t-1)}\left(\varphi^{R}(t-1,\omega,0)\right)
\succeq  Q
\gg
 0
\end{equation}
Thus the hypotheses of Theorem~\ref{LSD} are satisfied and
precisely one of the assertions~\textbf{a)} or~\textbf{b)} holds.
We show assertion~\textbf{a)} does not hold. Assume
that~\textbf{a)} holds on the contrary. Then, there exists
$P_{0}\in\mathbb{S}_{+}^{M}$ such that
\begin{equation}
\label{step3}
\mathbb{P}^{\mathcal{D}}\left(\lim_{t\rightarrow\infty}
\left\|\varphi^{R}\left(t,\theta^{R}_{-t}\omega,P_{0}\right)\right\|=
\infty\right)=1
\end{equation}
Then, for every $K\in\mathbb{T}_{+}$, we have
\begin{equation}
\label{step4}
\lim_{t\rightarrow\infty}\left\|\varphi^{R}
\left(t,\theta^{R}_{-t}\omega,P_{0}\right)\right\|>K,
\:\:\mathbb{P}^{\overline{\gamma}}\:\mbox{a.s.}
\end{equation}
It can be shown then (see~\cite{Riccati-weakconv} for details)
\begin{equation}
\label{step8}
\lim_{t\rightarrow\infty}\mathbb{P}^{\overline{\gamma}}
\left(\left\|\varphi^{R}\left(t,\theta^{R}_{-t}\omega,P_{0}\right)
\right\|>K\right)=1
\end{equation}
Since the above holds for every $K\in\mathbb{T}_{+}$, we have
\begin{equation}
\label{step9}\lim_{K\rightarrow\infty}\sup_{t\in\mathbb{T}_{+}}
\mathbb{P}^{\overline{\gamma}}\left(\left\|\varphi^{R}
\left(t,\theta^{R}_{-t}\omega,P_{0}\right)\right\|>K\right)=1
\end{equation}
On the other hand,
the stochastic boundedness of the RARE sequence and
Lemma~\ref{orbit_lemma} imply
\begin{eqnarray}
\label{step10} & &
\lim_{K\rightarrow\infty}\sup_{t\in\mathbb{T}_{+}}\mathbb{P}^{\mathcal{D}}
\left(\left\|\varphi^{R}\left(t,\theta^{R}_{-t}\omega,P_{0}\right)\right\|>K\right)\nonumber \\ &
= & \lim_{K\rightarrow\infty}\sup_{t\in\mathbb{T}_{+}}
\mathbb{P}^{\mathcal{D},P_{0}}\left(\left\|P_{t}\right\|>K\right)\nonumber
\\ & = & 0
\end{eqnarray}
This contradicts eqn.~(\ref{step9}) and~\textbf{a)} does not hold. Thus~\textbf{b)}
holds, and we have the result.
\end{proof}

Lemma~\ref{step} establishes the existence of a unique almost
equilibrium $u^{\mathcal{D}}$ if the schedule $\mathcal{D}$ is weakly detectable (or more generally leads to stochastic boundedness of the RARE sequence.) From the
distributional equivalence of pull-back and forward orbits, it
follows that the transition semigroup generated by the RARE Markov process is uniquely ergodic, i.e., the Markov process has a unique invariant probability. However, to show that
the measure induced by $u^{\mathcal{D}}$ on
$\mathbb{S}^{M}_{+}$ is attracting,\footnote{Attracting here refers to convergence in distribution to the unique invariant measure from every initial condition.}
eqn.~(\ref{step1}) must hold for all initial $v$.
Lemma~\ref{step} establishes convergence for a restricted class of
initial conditions $v$. We need the following result
to extend it to general initial conditions, whose proof follows from properties
of the Lyapunov and Riccati operators.

\begin{lemma}
\label{eq} For a schedule $\mathcal{D}$,
let $u^{\mathcal{D}}$ be an almost equilibrium of the RDS
$\left(\theta^{R},\varphi^{R}\right)$. Then
\begin{equation}
\label{eq1}
\mathbb{P}^{\mathcal{D}}\left(\omega:u^{\mathcal{D}}(\omega)\succeq
Q\right)=1
\end{equation}
\end{lemma}
%
%
\begin{proof}[Proof of Theorem~\ref{main_th}] We now complete the proof of Theorem~\ref{main_th}. We only highlight the key steps, the details follow the development in~\cite{Riccati-weakconv}. Let $\mu^{\mathcal{D}}$
be the distribution of the unique almost equilibrium in
Lemma~\ref{step}. By Lemma~\ref{eq} we have
$\mu^{\mathcal{D}}\left(\mathbb{S}_{++}^{M}\right)=1$.
Let $P_{0}\in\mathbb{S}_{+}^{M}$ be an arbitrary initial state.
Recall $\Omega^{\ast}$ as the $\theta^{R}$-invariant set with
$\mathbb{P}^{\overline{\gamma}}(\Omega^{\ast})=1$ in
Lemma~\ref{step} on which the almost equilibrium
$u^{\mathcal{D}}$ is defined. By Lemma~\ref{eq}, there
exists $\Omega_{1}\subset\Omega^{\ast}$ with
$\mathbb{P}^{\mathcal{D}}(\Omega_{1})=1$, such that
\begin{equation}
\label{main_res5} u^{\mathcal{D}}(\omega)\succeq
Q,\:\omega\in\Omega_{1}
\end{equation}
Define the random variable
$\widetilde{X}:\Omega\longmapsto\mathbb{S}_{+}^{M}$ by
\begin{equation}
\label{main_res6} \left\{ \begin{array}{ll}
                    P_{0} & \mbox{if $\omega\in\Omega_{1}$} \\
                    0 & \mbox{if $\omega\in\Omega_{1}^{c}$}
                   \end{array}
          \right.
\end{equation}
Now choose $\alpha>0$ sufficiently large, such that,
$P_{0}\preceq\alpha Q$, which is possible because $Q\gg 0$. Then it can be shown
\begin{equation}
\label{main_res8} 0\preceq\widetilde{X}(\omega)\preceq\alpha
u^{\overline{\gamma}}(\omega),\:\omega\in\Omega^{\ast}
\end{equation}
Then, by Lemma~\ref{step}, we have
\begin{equation}
\label{main_res11}
\lim_{t\rightarrow\infty}\varphi^{R}\left(t,\theta^{R}_{-t}\omega,\widetilde{X}
\left(\theta^{R}_{-t}\omega\right)\right)=
u^{\overline{\gamma}}(\omega),\:\omega\in\Omega^{\ast}
\end{equation}
which implies
convergence in distribution, i.e.,
\begin{equation}
\label{main_res12}
\varphi^{R}\left(t,\theta^{R}_{-t}\omega,\widetilde{X}
\left(\theta^{R}_{-t}\omega\right)\right)
\Longrightarrow\mu^{\overline{\gamma}}
\end{equation}
as $t\rightarrow\infty$. Then by
Lemma~\ref{orbit_lemma}, the sequence
$\left\{\varphi^{R}\left(t,\omega,\widetilde{X}(\omega)\right)
\right\}_{t\in\mathbb{T}_{+}}$
also converges weakly to the unique stationary
distribution $\mu^{\overline{\gamma}}$.

Now, since $\mathbb{P}^{\overline{\gamma}}(\Omega_{1})=1$, by
eqn.~(\ref{main_res6})
\begin{equation}
\label{main_res14}
\varphi^{R}\left(t,\omega,P_{0}\right)=
\varphi^{R}\left(t,\omega,\widetilde{X}(\omega)\right),
\:\mathbb{P}^{\overline{\gamma}}\:a.s.,\:t\in\mathbb{T}_{+}
\end{equation}
from which we can conclude that
\begin{equation}
\label{main_res16} P_{t}\Longrightarrow\mu^{\overline{\gamma}}
\end{equation}
as $t\rightarrow\infty$.
\end{proof}

\subsection{Proof outline for Theorem~\ref{supp_inv}}
\label{proof_supp_inv} The proof is lengthy and will be treated in detail elsewhere. The general line of arguments follow the corresponding in~\cite{Riccati-weakconv} for the single sensor case. It can be shown that the Markov process $\{P(t)\}$ possesses the weak Feller property, hence the support of its attracting probability is the intersection of the topological lower limits of the forward orbits emanating from all initial conditions (see, for example,~\cite{Zaharopol}.) The result can then be obtained by analyzing the limit properties of the maps $f_{i}$s. We refer the interested reader to~\cite{Riccati-weakconv}, which although deals with the single sensor problem, reflects the key technical arguments necessary for the development.

\section{Conclusions}
\label{conclusions}
We considered a broad class of estimation problems arising in networked control systems and established the ergodicity of the optimal estimation error process under fairly general conditions. The resulting invariant distribution is not absolutely continuous w.r.t. the Lebesgue measure in general, and we explicitly identify the support of its invariant distribution. We envision the applicability of the techniques developed in this paper to the pathwise analysis of more general hybrid or switched systems.

\section{Acknowledgements}
This work was partially supported by NSF under grants \#~ECS-0225449
and~\#~CNS-0428404.

\bibliographystyle{IEEEtran}
\bibliography{IEEEabrv,CentralBib}

\end{document}